\documentstyle[12pt,epsf]{article}

\author{
~\\
Tsvi PIRAN\\
Racah Institute for Physics\\
The Hebrew University, Jerusalem 91904, Israel\\
~\\
}
\title{On Gravitational Repulsion}
\def\etal{{\it et. al., }}

\def\bootes{Bo\"otes}
\def\potent{{\textsc{potent}}}

\def\vf{{\textsc{void finder}}}
\def\h{h^{-1}}
\def\2d{two-dimensional}
\def\3d{three-dimensional}
\def\apj{\emph{Ap. J.}}
\def\apjl{\emph{Ap. J. Lett.}}

\def\mnras{\emph{MNRAS}}
\def\araa{\emph{Ann. Rev. Astron. Astrophys.}}

\def\sci{\emph{Science}}

\begin{document}            

\maketitle

%\newpage
\begin{abstract}
The concepts of negative gravitational mass and gravitational
repulsion are alien to general relativity. Still, we show here that
small negative fluctuations - small dimples in the primordial density
field - that act as if they have an effective negative gravitational
mass, play a dominant role in shaping our Universe.  These initially
tiny perturbations repel matter surrounding them, expand and grow to
become voids in the galaxy distribution. These voids -  regions with a
diameter of $40 \h$ Mpc which are almost devoid of galaxies - are the
largest object in the Universe.
\end{abstract}

\bigskip
{\bf 1. Strange and Not So Strange Negative Masses}
\bigskip

Two (gravitational) negative masses attract each other.  The
equivalence principle requires that their inertial masses  is
negative as well. In this case the masses  move away from each
other.  A negative and positive mass pair is even stranger.  They
repel each other.  The positive mass moves away from the negative one
but the negative mass accelerates towards the positive one. The
distance between the masses does not vary while they are jettisoned
together at a constant acceleration.  Momentum and energy are conserved
as the negative momentum and kinetic energy of the negative mass
balance the positive energy and momentum of the positive one.

This is so strange that in spite of the equivalence principle it is 
worthwhile to consider particles with a negative gravitational mass
and a positive inertial mass. These are somewhat more reasonable.  Two
such masses attract each other and as a result they move towards each
other. A (gravitational) negative mass repels a (gravitational)
positive mass and now both masses move away from each other. This
resembles, of course, electrically charged particles (with the
corresponding change in the sign of the force).  A test particle with
an ``effective'' negative gravitational mass and a positive inertial
mass can be easily simulated. A light object immersed in a denser
fluid feels an ``effective'' repulsive gravitational force that mimics
the gravitational repulsion felt by a test particle with a negative
gravitational mass in the gravitational field of a positive mass.

This trivial example of a test particle raises the natural question.
Are there sources of ``effective'' repulsive gravitational force?  Are
there objects with an ``effective'' negative gravitational mass?
Surprisingly, the answer is yes. Moreover, these ``effective''
negative masses are responsible for the creation of the largest
structures in the Universe!

\bigskip
{\bf 2. Cosmological Negative Gravitational Masses}
\bigskip

We live in an expanding universe which is close to its critical
density (for simplicity we consider here an $\Omega = 1$
Universe). The Cosmic Microwave Background Radiation (CMBR) tells us
that in the past the Universe was practically homogeneous.  Our
existence, the observations galaxies and even the CMBR itself
\cite{Smoot} reveals  that the perfectly homogeneous FRW model is an
idealization.  Small primordial deviations, of order $10^{-5}$ at
horizon crossing, were present. These fluctuations grew during the
matter dominated era to become the very large deviations from
homogeneity observed today.

We understand very well how does a cosmological positive density
fluctuation evolve. This is not surprising, after all we live within
one!  A spherical overdense region behaves like a closed universe
within an outer flat one. This closed universe expands until it
reaches its maximal size and then it begins to
collapses\footnote{Somewhat strangely this phase is often referred to
as ``recollapse'' - however this region has never collapsed
before!}. The collapse continues until ``virialization'' when the
kinetic and gravitational energies are equal and the region has shrank
to half of its maximal physical size. This simple picture could be
misleading at times. Gravitational collapse is unstable to
non-spherical modes. Even if the spherical approximation is initially
valid it will break down during the collapse and flat pancakes rather
then round balls are more likely to form.  Most observed cosmological
objects, like galaxies or clusters, are indeed far from spherical.

Clearly there is an equal number of underdense and overdense
regions. Still the fate of underdense regions was, somehow, largely
ignored, and when it was discussed it was forgotten.  We show that
cosmological underdense regions behave like ``effective'' negative
gravitational masses (with a positive inertial mass). They repel
nearby positive masses and attract other underdense regions.  Using
this analogy we examine the evolution of cosmological underdense
regions and we find that primordial negative density perturbations,
small initial dimples in the density field, grow and become the
observed voids in the galaxy distribution.  Using recent galaxy
redshift surveys we demonstrate that these voids are the largest
objects in the Universe containing most of the volume of the Universe
today.

Consider an idealized spherical underdense region.  At some initial
time $t_i$ (e.g. at horizon crossing) it is characterized by a
comoving size $R_i$ and a negative fractional underdensity $\delta
\rho/\rho = -\epsilon_i$.  As long as the physical size is 
larger than the horizon the dimple is frozen with a constant
underdensity and a constant comoving size\footnote{This statement
depends, of course, on gauge choice - but for practical purposes it
yields a good description.}.  Once it crosses the horizon, the dimple
behaves, in analogy with an overdense region, like an open universe
within the outer flat universe. As an open universe it expands faster
than the surrounding flat one. Matter within it expands away from the
center of the dimple faster than the surrounding matter. This open
universe also influences strongly the surrounding regions.  A sphere
surrounding the dimple contains less mass than an equivalent sphere
elsewhere.  Consequently a shell on its boundary expands faster than
average, as if there is a negative mass repelling it.  The dimple has
an ``effective'' negative gravitational mass!  The surrounding matter
forms a high density ridge along the rim of the dimple.  At a time
$t_{sc}$ given by $\epsilon_i (t_{sc}/t_i)^{2/3} \approx
(1+z_i)/(1+z_{sc}) \approx \delta_{crit}$, (where $\delta_{crit}$
ranges between 2.5 and 4.5 depending on the initial velocity
distribution) the shell located just outside the dimple, which feels
the strongest repulsion overtakes its outer neighboring shell
\cite{blu92}. Shell crossing occurs. The density of the rim becomes
infinite and the model of an open universe within a flat one breaks
down. At $t_{sc}$ the comoving radius of the dimple $R_{sc}$ is 1.7
times the initial comoving radius $R_{i}$. The local density is then
$\approx 0.2$ of the average density.

After that a second phase begins in which shell crossing continues and
the expansion of the underdense region settles quickly to a self
similar solution \cite{Suto,fg84,ber85}.  The surrounding high density
ridge expands, in comoving coordinates, as $t^{2/9}\propto
(1+z)^{-1/3}$ and its comoving radius satisfies: $R / R_{sc} = \left
((1 +z_{sc}) /(1+z)\right )^{1/3}$. This expansion is much slower than
the expansion during the earlier phase.

Pairs or more complicated systems of multiple dimples are more
difficult to analyze. Generally they do not preserve their shape and
mass as matter flows out from them. Consequently even the simple ``two
body problem'' of two nearby underdense regions can be addressed only
using numerical N-body simulations. Dubinski \etal \cite{du93} have
considered several idealized configurations of interacting dimples.
They found that two nearby dimples expand towards each other repelling
the surrounding matter and creating a high density ridge between them.
This ridge is later broken, its matter is repelled outwards in a
direction perpendicular to the line connecting the two centers. At
this stage the two underdense regions merge to a single one.  This
region continues to expand outwards repelling surrounding matter and
becoming more and more spherical.  This  is a nice feature of 
underdense regions, making their analysis simpler. While a collapsing
region is unstable to non-spherical perturbations, an expanding one
becomes more spherical with time.

\bigskip
{\bf 3. Voids in the Galaxy Distribution}
\bigskip

Perhaps one of the most intriguing findings of dense and complete
nearby redshift surveys has been the discovery of numerous large voids
on scales of $\sim 50 \h$ Mpc in the galaxy distribution.  Although
the voids are a fundamental element of the large-scale structure (LSS)
of the universe, the realization that they dominate the LSS is
relatively recent \cite{gh89}.  Early surveys published during the
70's, like the Coma/A1367 redshift survey \cite{gt78} and the
Hercules/A2199 redshift survey \cite{crt81}, gave the first
indications for the existence of voids, each revealing a void with a
characteristic diameter of $\sim 20 \h$ Mpc.  Surprising as these
findings might have been, it was not before the discovery of the
\bootes{} void \cite{kir81} that the voids caught the attention of 
the astrophysical community (for a review about the early void
explorations, see \cite{rood88}).

The unexpectedly large void found in the \bootes{} constellation,
confirmed to have a diameter of $\sim 60 \h$ Mpc \cite{kir87},
naturally brought up the question whether the empty regions we observe
are a common feature of the galaxy distribution, or rather rare
exceptions.  Wide-angle yet dense surveys probing relatively large
volumes of the nearby universe, established that the voids are indeed
a common feature of the LSS, and as such must be incorporated into any
valid model of it. The  first slice from the
\emph{Center for Astrophysics} (CfA) redshift survey \cite{lap86}
revealed the picture of a universe where the galaxies are located on
the surfaces of bubble-like structures, with diameters in the range
$25-50 \h$ Mpc.  The extensions of the CfA survey \cite{gh89},
complemented in the south hemisphere by the \emph{Southern Sky
Redshift Survey} (SSRS) and its extension, the SSRS2 \cite{dc88,dc94}
have shown that not only large voids exist, but more importantly --
that they occur frequently (at least judging by eye), suggesting a
compact network of voids filling the entire volume.

Until recently no suitable algorithm was available to quantitatively
study the properties of voids.  Only some gross estimates were
inferred from visual inspection of the existing redshift
surveys. During the last two years we have developed the
\vf~~algorithm \cite{epd96,epd97,ep97} which identifies  voids in redshift
surveys and measures their size and underdensity. We have used the
\vf~~to analyze the void distribution in the SSRS2 survey \cite{epd96}
and the IRAS survey \cite{epd97} and we have verified the visual
picture of a void filled universe in which galaxies are mostly located
along walls surrounding voids (see Figs. 1 and 2).  In both surveys we
find that voids whose diameter is $\approx 40 Mpc \h$ with a typical
density of galaxies of order 10\% of the average density contain more
than half of the volume of the universe. These voids are clearly the
largest object observed in the Universe.

%\newpage
\bigskip
{\bf 4. Voids, Negative Masses  and the LSS}
\bigskip

The observed correlation between voids in the galaxy distribution
found by the \vf~~ \cite{epd97} and regions with low dark matter
density found by \potent~~ \cite{dekel90} show clearly that the origin
of the voids must be gravitational. The previous analysis of the
evolution of an underdense regions suggests the following picture:
Primordial underdense regions - dimples - act like cosmological
``negative'' masses. These dimples are the seeds of the observed
voids. While overdense regions collect more and more matter and shrink
in both real and comoving sizes, dimples repel matter and expand. We
can view the centers of the underdense regions as effective
``negative'' gravitational masses that repel matter. The repelled
matter is aligned along walls located between the ``negative''
centers. Voids, with low galaxy and dark matter densities are centered
on these ``negative'' masses and are surrounded by walls.  Eventually
the walls are torn apart, the voids merge and a network of larger
voids on a large scale forms.

It is illuminating to consider underdense regions of a given comoving
scale $\lambda$.  These dimples cross the horizon at the moment when
their physical size equals horizon's size, with a typical amplitude
$\epsilon_i(\lambda)$ (which is scale independent if the primordial
spectrum is a scale independent one).  The dimples grow in amplitude
and in comoving size.  By the time that they reach shell crossing
their comoving size has increased by a factor of 1.7.  With a typical
density of 20\% of the average density they definitely qualify as
voids.  These voids practically touch each other and fill the
universe. Later the walls between the voids break down. The voids
merge and new voids on a larger scale appear. The network of voids is
replaced in an effective self-similar manner by a network of voids
with a larger characteristic scale. At each moment dimples that reach
shell crossing form the current prominent voids, these are destroyed
latter forming larger voids and so on.

Denoting the current radius of the observed voids by $R_{voids}$ we
find $\lambda_{voids}=R_{voids}/1.7$ and $\epsilon_i (\lambda_{voids})
[1+z_i(\lambda_{voids})] =\delta_{crit}$. Amazingly $\epsilon_i$
determined in this way approximately equals $ 10^{-5}$, as determined
from extrapolation of CMBR observations
\cite{Smoot}\footnote{A detailed analysis \cite{pi93} shows that 
this estimate indicates a slightly larger primordial fluctuations than
what other methods, such as a scale independent interpretation of the
CMBR data or observations of rich clusters, yield.  However, one can
think of numerous refinements of both types of estimates.}.  This
overall agreement is impressive.  It demonstrates that gravitational
repulsion, caused by dimples in the primordial density field -
cosmological ``negative'' masses - create the voids, the largest
structures in the Universe.

%I thank Hagai El-Ad, Luiz Da Costa, and Dalia Goldwirth for many
%helpful discussions concerning dimples and their daughthers the voids.

\newpage
\begin{figure}
\centering
\epsfbox{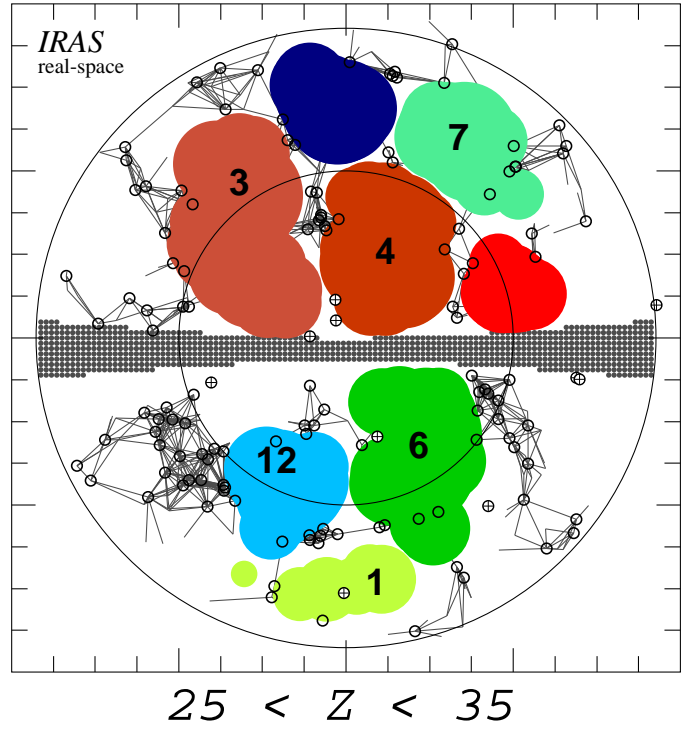}
\caption{ Voids in the \textit{IRAS} 1.2-Jy survey. The colored areas mark the
intersection of the $ \mbox{SGZ} = 30\:h^{-1}\:\mbox{Mpc} $ plane with
the three-dimensional voids. Void~4 is the Local Void. The walls
surrounding the voids are highlighted by drawing dark lines connecting
nearby wall-galaxies. The dark area marks the ZOA, caused by the Galaxy.}
\end{figure}
\newpage
\begin{figure}
\centering
\epsfbox{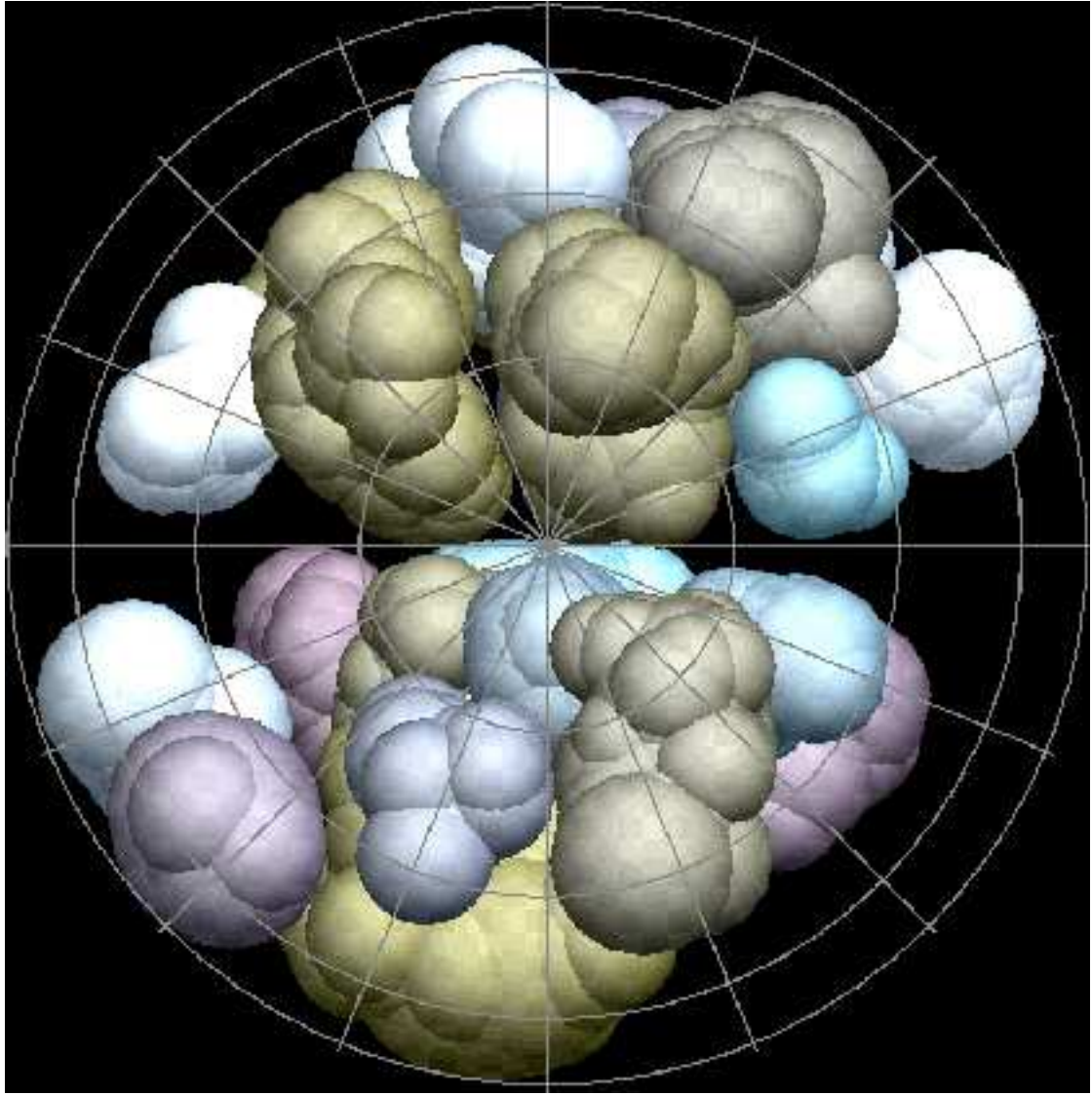}
\caption{ Three-dimensional view of the voids in the
\textit{IRAS} survey. The ZOA, caused by the Galaxy, runs horizontally
across the image. The area at the left, near the ZOA, with no voids,
corresponds to the Great Attractor. The absence of voids from the
lower, right-hand part of the image, is due to the Cetus wall and the
Perseus-Pisces supercluster.}
\end{figure}
\end{document}